\documentclass[aps,pra,manuscript,groupedaddress]{revtex4-1}

\usepackage[utf8]{inputenc}
\usepackage[T1]{fontenc}

\usepackage{amssymb}

\usepackage{amsmath}

\usepackage{nicefrac}

\usepackage{units}

\usepackage{esint}

\usepackage{braket}

\usepackage{color}

\usepackage{graphicx}

\begin{document}

\title{Classical fields and quantum measurement \\ for Bose-Einstein condensate}

\author{Tomasz G\'orski\textsuperscript{1,2}}
\author{Kazimerz Rz\k{a}\.zewski\textsuperscript{1,2} }

\affiliation{
\textsuperscript{1} Center for Theoretical Physics, Polish Academy of Sciences, Al. Lotnik\'ow 32/46, 02-668 Warsaw, Poland \\
\textsuperscript{2} 5. Physikalisches Institute, Universit\"at Stuttgart, Pfaffenwaldring 57, 70550 Stuttgart, Germany}

\date{\today}

\begin{abstract}
We analyze a process of splitting of the Bose-Einstein condensate and the mutual coherence of two separated atomic clouds. Within the classical fields approximation we show that coherence between clouds is degraded if atoms interact and if we account for the sufficiently long observation time. We also show, that upon recombination, the coherence across the sample is restored. The coherence is not fully degraded if the splitting potential remains sufficiently penetrable. We calculate the variance of atom number difference for this time-averaging measurement and show that for low temperatures it can be well below Poissonian limit like it was observed in the experiments. 
\end{abstract}


\maketitle

\section{Introduction}
Although a long time has passed since the early study of the quantum measurement of John von Neumann to a modern analysis of Wojciech \.Zurek \cite{zurek81, zurek00, zurek12} stressing the role of decoherence due to an interaction of a quantum system with the macroscopic, classical measuring device, a lot remains to be done before the understanding of the quantum measurement could be considered complete.

Experiments with a Bose-Einstein condensate bring a new element to the game. The condensate is a mesoscopic collection of very many atoms, nearly perfectly isolated from any contact with the environment, forming a well defined single quantum object. In the experiments with BEC we are probing a border between a macroscopic, classical world and the quantum realm in yet another way. In most experiments, the measurement of the condensate is done by shining a laser pulse on the gas and then analyzing the shadow with the help of a ccd camera. The pulses have their length, typically several microseconds, and the pixels of the ccd camera have their extend allowing these days for a spatial resolution below 1 micron. Thus the procedure introduces inevitable limits on the temporal and the spatial resolution. A natural question arises if these limits influence the results of a measurement in a significant way.

In the present paper we argue that it does, at least within a very convenient approximate treatment of the weakly interacting Bose gas known as the classical fields approximation. Our result raises a question if the same is true in the experimental environment. In this paper we concentrate our attention on experimentally relevant, yet very fundamental process of splitting of a condensate by a dividing potential. This process is an essential ingredient in many atomic interference schemes \cite{shin04, schumm05, jo07}. The coherence properties induced by the splitting and recombination were studied for instance in \cite{mebrahtu06}.

Let us consider the simplest possibility: the ground state of a weakly interacting Bose gas confined in a trap. In this case the multiparticle wave function, to a very good approximation, is a product of all atoms occupying the same normalized orbital $\phi(x)$:
\begin{equation} \label{eq.prod_state}
\Psi(x_1, x_2, \dots x_N) = \prod_{i=1}^{N} \phi(x_i)
\end{equation} 

Now, we consider two spatial parts of this system labeled $L$ and $R$ respectively and ask what is the probability distribution of finding $N_L$ atoms in the $L$ part (and remaining $N - N_L$ in the $R$ part).
Each orbital may be written as a sum of two mutually orthogonal orbitals
$\phi_L$ and $\phi_R$ with the support in $L$ and $R$ respectively:
\begin{equation}
\phi(x) =\sqrt{p} \phi_L (x) + \sqrt{1 - p} \phi_R (x)
\end{equation}
where
\begin{equation}
p = \int_L \left | \phi(x) \right |^2
\end{equation}
This combinatorial problem has a very simple solution. The probability distribution of finding $N_L$ atoms in the left part is binomial
\begin{equation}
p(N_L) = \binom{N}{N_L} p^{N_L} (1 - p)^{N - N_L}
\end{equation}
This implies that the mean value of the population difference $\Delta N = N_L - N_R = 2N_L - N$, a quantity most relevant in atom interferometry \cite{shin04, schumm05}, is equal to
\begin{equation}
\braket{\Delta N} = N(2p -1)
\end{equation}
and its variance
\begin{equation} \label{eq.var}
\mathrm{Var}(\Delta N) = 4Np(1-p)
\end{equation} 

Thus the variance is vanishing for the whole $\phi$ either in the left or in the right part of the trap ($p = 0$ or $p = 1$) and is maximal for symmetric splitting ($p = 1/ 2$) indicating Poissonian fluctuations in this case.

In several experiments the trapped Bose gas was physically split into two parts using a dividing potential produced by a laser beam or the magnetic potential barrier on an atom chip \cite{maussang10}. Then the statistics of the population difference was measured as a function of initial temperature. As the temperature is lowered and thermal fluctuations are being suppressed, the population difference becomes clearly sub-Poissonian Fig. 4 in \cite{maussang10}. To account for this observation in \cite{maussang10} it was assumed that the system after splitting becomes a classical mixed state of left and right modes. The coherence between the two parts is lost. On the other hand very early experiments with the recombination of the split condensate showed mutual coherence of the two parts by the observation of interference fringes \cite{ketterle97}.

This looks like a paradox. It is the main purpose of this paper to demonstrate that both: the loss of coherence as a result of splitting and its revival at recombination arise in a natural way in the classical fields approximation for weakly interacting bosons \cite{kagan97, goral01, sinatra01, davis01} and that the coarse graining of an imperfect measurement is essential to explain both effects.

In the classical fields approximation the atomic field operator $\hat{\Psi}(x)$, which is conveniently expanded into the empty trap eigenvector basis:
\begin{equation}
\hat{\Psi}(x) = \sum_i \phi_i (x) \hat{a}_i
\end{equation}
($\phi_i(x)$ harmonic oscillator eigenstates for the quadratic trapping potential) is being replaced by a c-number classical, complex, scalar field with complex amplitudes $\alpha_j$ substituted in place of the annihilation operators $\hat{a}_j$:
\begin{equation}
\hat{\Psi}(x) \rightarrow \Psi(x) = \sum_{i=0}^{K_{max}} \phi_i (x) \alpha_i
\end{equation} 
The relation between the atomic field operator and its classical field simplification is analogous to the relation between Maxwell electric and magnetic fields and their operator counterparts in QED. Just as for light, the classical approximation suffers for an ultraviolet catastrophe when thermal equilibrium is considered, thus a high energy cut-off $K_{max}$ is needed. Its optimal choice was determined in \cite{witkowska09}. For a 1D Bose gas with contact repulsive interactions in a harmonic trap of frequency $\omega$ at temperature $T$ it is given by:
\begin{equation}
K_{max} = \frac{k_B T + \mu}{\hbar \omega}
\end{equation}
where $\mu$ is the chemical potential and $k_B$ is the Boltzmann constant. Now, the system lives in a finite dimensional classical phase space spanned by the amplitudes $\alpha_j$. The canonical probability distribution in this phase space is given by:
\begin{equation} \label{eq.prob_dist} 
P(\{ \alpha \}) = \frac{1}{Z} \exp (-\frac{E(\{ \alpha \})}{k_B T}
\end{equation}
where $Z$ is the canonical partition function while the energy of a given point in the phase space
is:
\begin{equation}
E(\{ \alpha \}) = \hbar \omega \sum_{j} j \left |  \alpha_j \right |^2 + E_{int}(\{ \alpha \})
\end{equation}
The first term represents the energy of an ideal gas in a harmonic trap and the second term is the
interaction energy -- a quartic polynomial in the amplitudes, which for the contact interaction $V(x - y) = g \delta(x - y)$ is simply:
\begin{equation}
E_{int}(\{ \alpha \}) = \frac{g}{2} \int \left | \Psi(x) \right |^4 \mathrm{d}x
\end{equation}
The distribution (\ref{eq.prob_dist}) may be conveniently generated by the Metropolis algorithm \cite{metropolis53}. We interpret individual elements of the ensemble as describing individual runs of the experiment. This way, for instance, spontaneously generated solitons in quasi 1D Bose gas \cite{karpiuk12} were discovered. Obviously as the temperature grows, the fields of the ensemble are typically not perfectly symmetric even for symmetric splitting. The fluctuation of the population difference due to this effect we will call “thermal”. However, there is an additional question: To what extend the fluctuations associated with the multiparticle occupation of the same quantum state (here we will call them “quantum fluctuations”) of the kind given by (\ref{eq.var}) contribute to the measured population difference variance.
We stress that the classical field for nonzero temperature should not be confused with the universal single particle wave function of the ground state~(\ref{eq.prod_state}).

In Ref. \cite{goral02} it was argued and then numerously exploited \cite{brewczyk13}, that the measured one-particle density matrix may be obtained from the classical fields by means of the coarse graining reflecting the limited spatial and/or temporal resolution of available detectors. Thus, physical results may be obtained from: 
\begin{equation} \label{eq.mean}
\rho(x,x',t) = \frac{1}{\Delta x \Delta x' \Delta \tau} \iint_{\Delta x \Delta x' \Delta \tau} \Psi^{\ast} (x + \zeta, t + \tau) \Psi (x' + \zeta, t + \tau) \mathrm{d} \zeta \mathrm{d} \tau
\end{equation}

Again a close analogy exists with the classical, Maxwell electrodynamics. Imagine a source of a partially coherent light. At a microscopic level at each point in space and time electric field has a well defined value. While a product of electric fields at two points in space-time is a well defined but usually useless number, a coarse graining caused by the detectors makes such a quantity useful. This way one defines such notions like the coherence length and the coherence time. Only such a coarse grained quantity may be checked against various stochastic models of partially coherent light.

The action of the coarse graining process may be illustrated by a simple example of the ideal gas.
Let us assume that the classical field for some atomic sample is a sum of the eigenstates  $\psi_j$ of its single
particle Hamiltonian with energies $E_j$ that is $\Psi(x,t) = \sum_j \alpha_j \exp(-i E_j t) \psi_j (x)$, then for time integration
much longer than typical inverse of a distance between the energies, the relative phases between
different amplitudes are forgotten and the coarse grained one-particle density matrix reads: $\rho(x,x') = \sum_j |\alpha_j|^2 \psi_j^{\ast}(x) \psi_j (x')$.

Like in the classical theory of coherence of light \cite{mandel95}, also for the atomic correlation function (\ref{eq.mean}), viewed as a hermitian operator in positions $x$ and $y$, may be diagonalized
\begin{equation}
\rho(x, x') = \sum_i \lambda_i \phi_{i}^{\ast} (x) \phi_{i}(x') 
\end{equation}
The eigenvectors, known in the coherence theory as the coherence modes (in atomic physics known as natural orbitals \cite{lodwin55}) are just the wave functions occupied by the atoms. The eigenvalues tell their respective populations \cite{penrose56}
\begin{equation}
\lambda_i = \frac{N_i}{N}
\end{equation}
The binomial fluctuations in each orbital are mutually independent, thus the quantum fluctuations of atom number difference is given as a sum of these independent contributions:
\begin{equation}  \label{quantum_variance}
\mathrm{Var}(\Delta N) = 4 N \sum_i \lambda_i p_i (1 - p_i)
\end{equation}

\section{Method and results}

Equipped with this theoretical background we turn now to an illustrative example: We consider a one dimensional weakly interacting Bose gas confined in a harmonic trap of frequency $\omega  = 29 \mathrm{Hz}$. The coupling constant $g$ is equal to $0.03$ and is calculated assuming that the 1D treatment is an approximation to an elongated axially symmetric 3D harmonic trap with transverse frequency $\omega_\perp = 1 \mathrm{kHz}$, $g = \frac{1}{2\pi} \frac{m\omega_\perp}{\hbar} g_{3D}$.

At first we prepare our system of $N = 900$ \textsuperscript{87}Rb atoms. A single point in the phase space is then used as an initial state for the process of growing dividing potential of Gaussian shape
$V_d(x,t) = A(t) e^{-\frac{x^2}{\sigma^2}}$.
Linear ramp $A(t) = Rt$ is assumed with the rate $R = 100 \hbar \omega^2$. The time evolution of the classical field is governed by the time dependent Gross-Pitaevskii equation:
\begin{equation}
i \hbar \frac{\partial \Psi}{\partial t} = \left [ \frac{p^2}{2m} + \frac{m \omega^2}{2} x^2 + V_d (x, t) + g \left | \Psi (x, t) \right |^2 \right ]  \Psi (x, t)
\end{equation}
The growth terminates at $A = 200 \hbar \omega$, when tunnelling across the barrier is no longer possible. We then perform the coarse graining using different exposure times $\Delta \tau$.

\begin{figure}[h]
\includegraphics[width=1\textwidth]{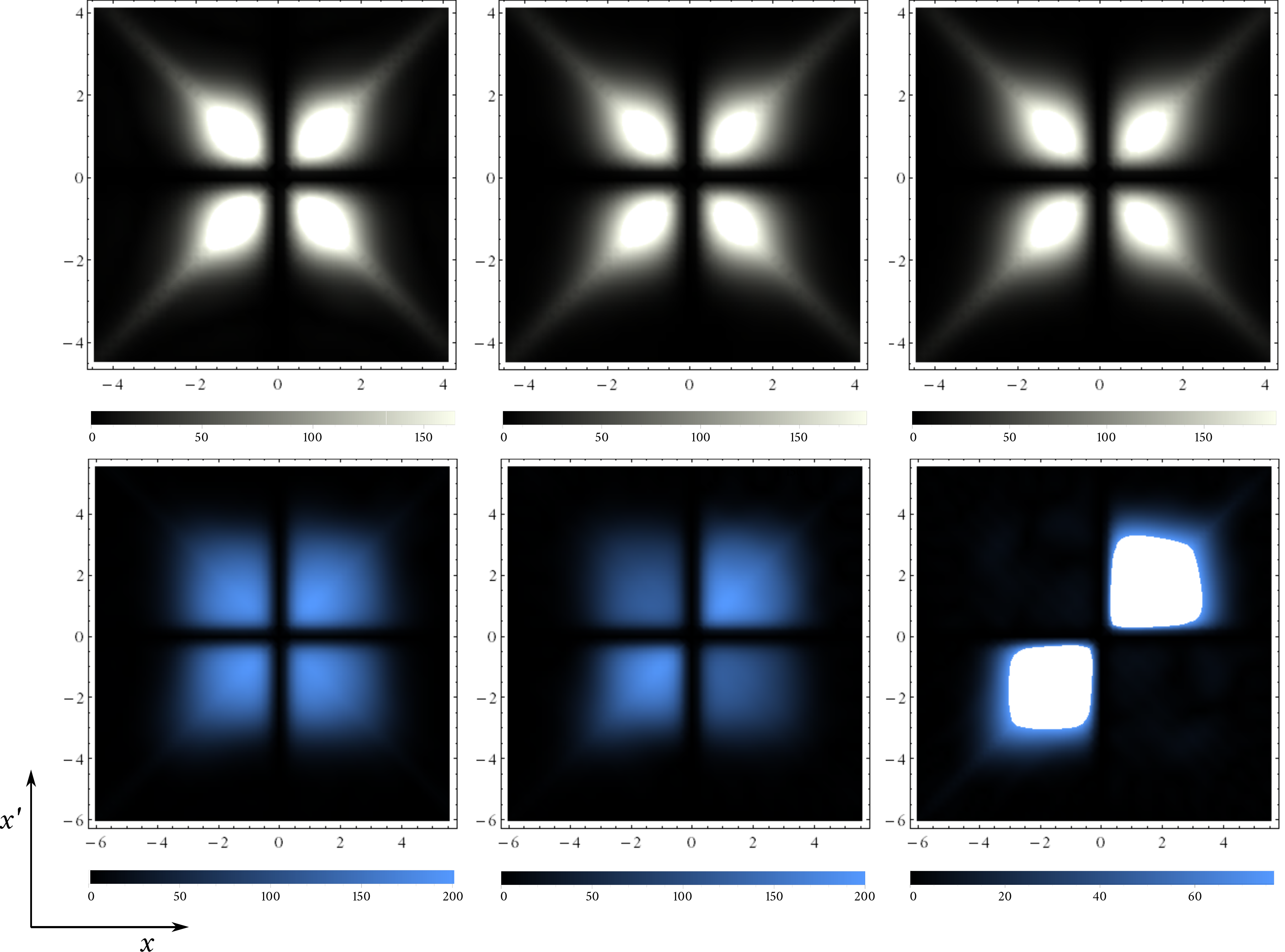}

\caption{The coarse grained density matrices after the splitting for three averaging times 0.34 s, 0.69 s, 1.38 s -- for the ideal gas (upper row) and the interacting gas (lower row) at temperature $0.08 \ T_c$. The squared modulus of a density matrix is plotted here and in subsequent figures.}
\label{dm_INT+ID_T5}
\end{figure}

We compare the results for the ideal Bose gas with that for the weakly interacting one. For the ideal gas the mutual coherence between the two parts is not degraded by even a very long observation time.
On the contrary for the interacting gas at low temperature ($0.08 T_c$, where $T_c = N/\mathrm{ln}(N)$ is a transition temperature for 1D gas) the phase of the field in each well evolves at nearly uniform rate which depends on the chemical potential, i.e. the number of atoms in each well.
Therefore the elements of coarse grained density matrix in off-diagonal quadrants oscillate and decay slowly with time.
The resulting coarse-grained density matrices are shown in Fig. \ref{dm_INT+ID_T5}.

\begin{figure}[h]
\includegraphics[width=1\textwidth]{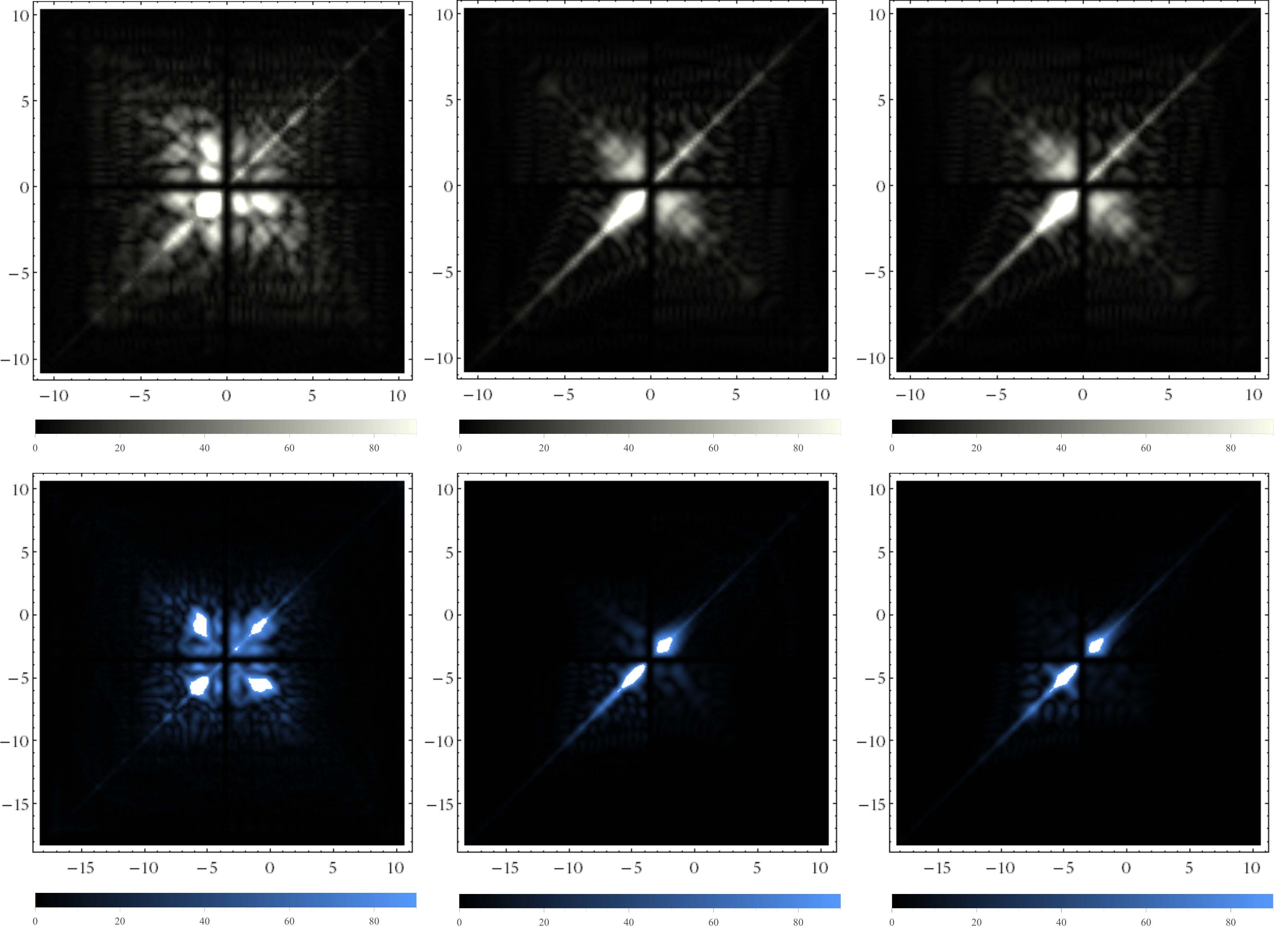}

\caption{The coarse grained density matrices after the splitting for three averaging times 0.34 s, 0.69 s, 1.38 s -- for the ideal gas (upper row) and the interacting gas (lower row) at temperature $0.76 \ T_c$.}
\label{dm_INT+ID_T100}
\end{figure}

However for high temperature ($T = 0.76 T_c$) we do not observe any oscillations: the decay in the off-diagonal quadrants is monotonic. When the observation time is long enough the state is almost an incoherent mixture of the left and right components. The resulting coarse-grained density matrices are shown in Fig. \ref{dm_INT+ID_T100}.

To explain the behaviour of a coarse grained density matrix at low temperature we introduce a naive two mode model in which a state vector has two components -- each corresponding to a state localized in the left or the right well. Time evolution of this state is given by
\begin{equation}
\ket{\sqrt{p_L} e^{-i\mu_L t}, \sqrt{p_R} e^{-i\mu_R t}}
\end{equation}
where $p_L$ and $p_R$ are the probabilities of finding atom in the left an in the right well respectively. We calculate the chemical potentials $\mu_L, \mu_R$ analyzing the GP time evolution of ground states in the left and the right well. Subsequently we can calculate the time evolution of the coarse grained density matrix. We return to this model below.

\begin{figure}
\includegraphics[width = 1 \textwidth]{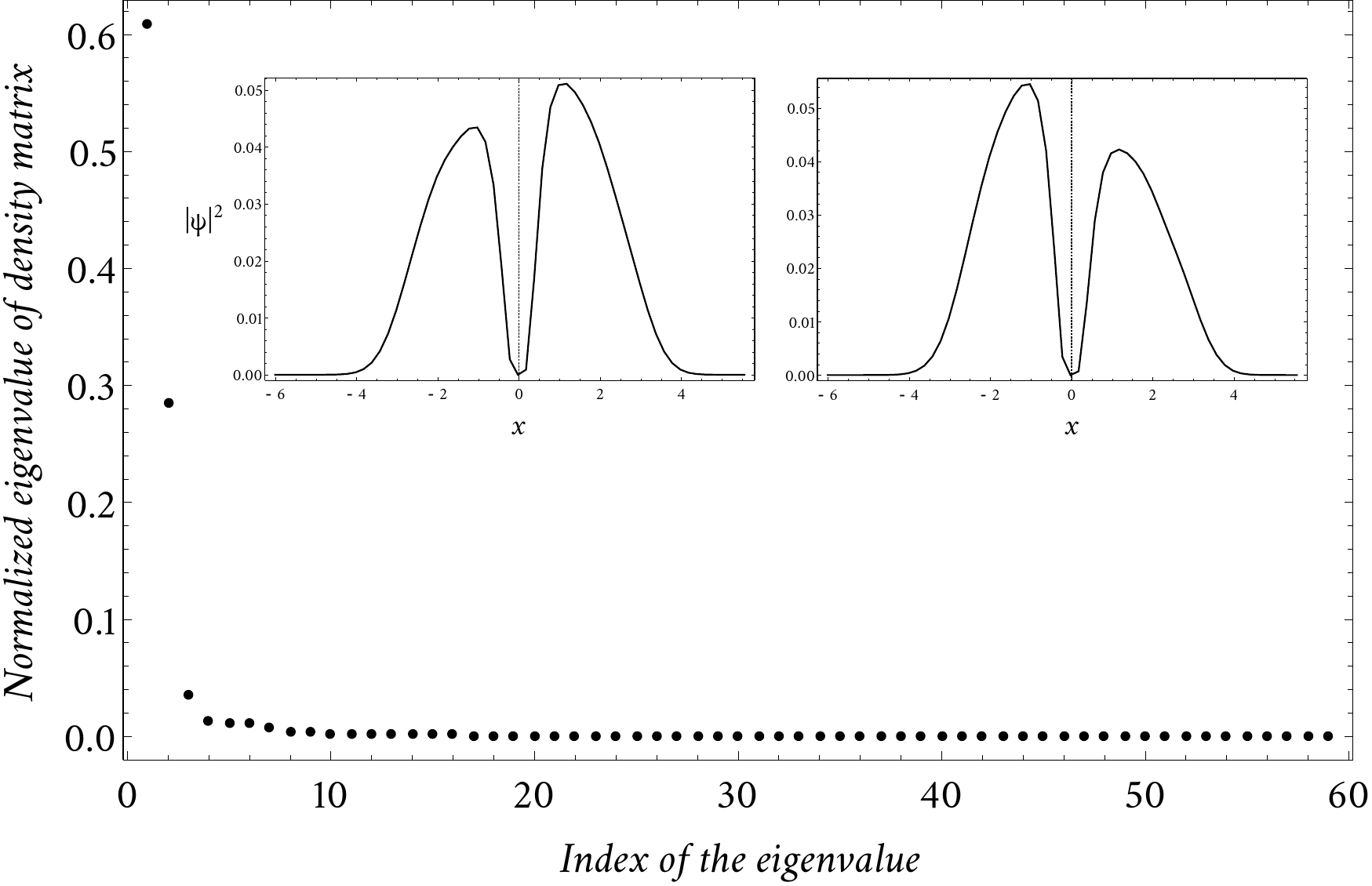}
\caption{The eigenvalues of the coarse grained density matrix and the density profiles of two dominant coherence modes (in the inset) after complete splitting. The time of coarse graining is about 1.03 s. }
\label{evl+evc_tc30}
\end{figure}

\begin{figure}
\includegraphics[width = 1 \textwidth]{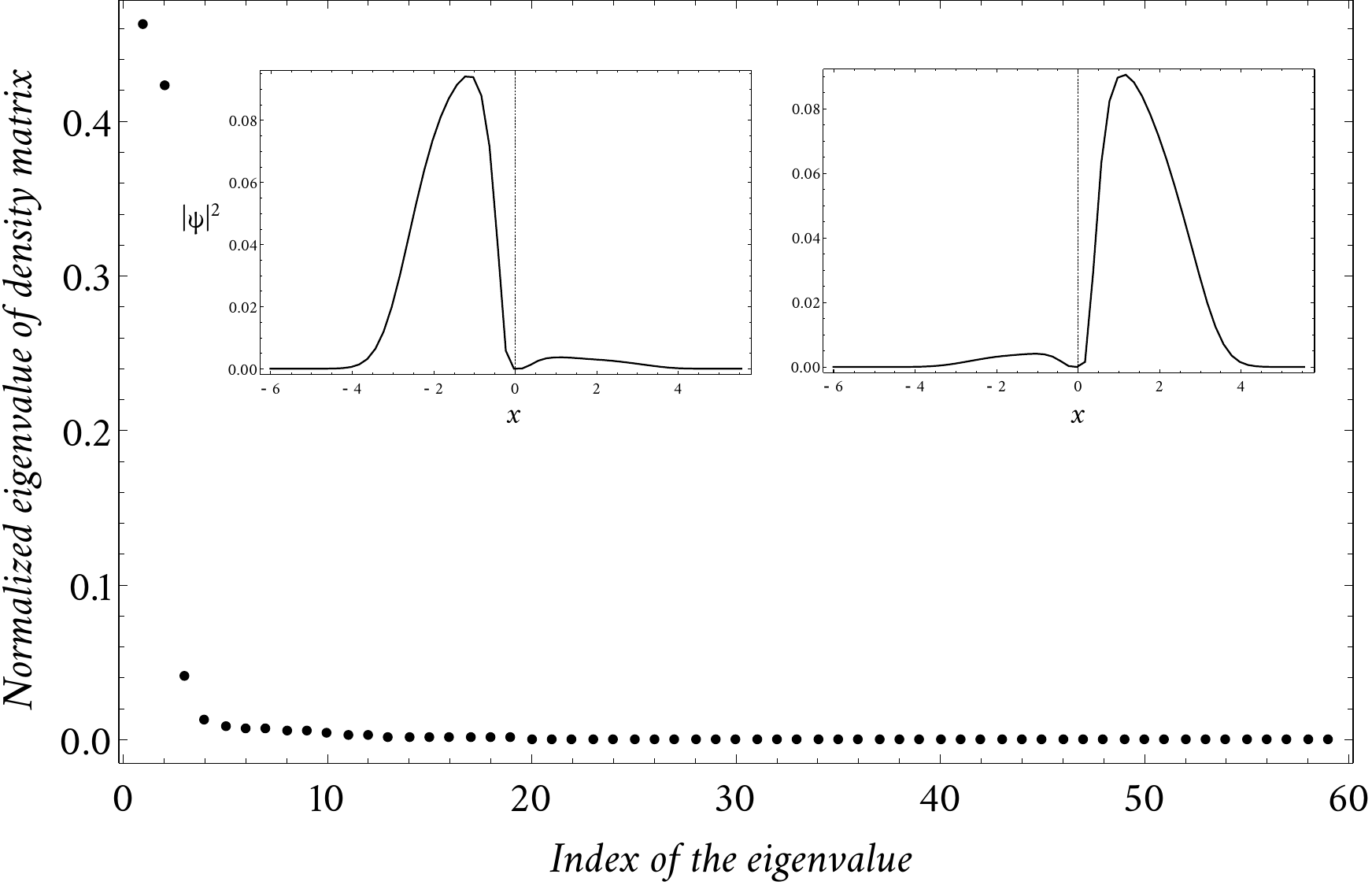}
\caption{The eigenvalues of the coarse grained density matrix and the density profiles of two dominant coherence modes (in the inset) after complete splitting. The time of coarse graining is about 1.38 s. }
\label{evl+evc_tc40}
\end{figure}

We also diagonalize the coarse grained density matrix. The distribution of the eigenvalues of the initial state has a dominant element -- the population of the condensed state \cite{penrose56}. This distributions are shown in Fig. \ref{evl+evc_tc30} and \ref{evl+evc_tc40}. The density matrix of nearly completely incoherent mixture of the left and the right subsystem has two dominant eigenvalues with the corresponding eigenvectors residing nearly fully in the left and in the right well respectively.

Thus, while at low temperature most atoms occupy the two lowest orbitals, each of them is so asymmetric, i.e. it has the value of $p$ so close to $0$ or to $1$, that the quantum fluctuations are negligible. This explains why in \cite{maussang10} a sub-Poissonian fluctuations of the population difference was observed at low temperature.

\begin{figure}[h]
\includegraphics[width=0.8\textwidth]{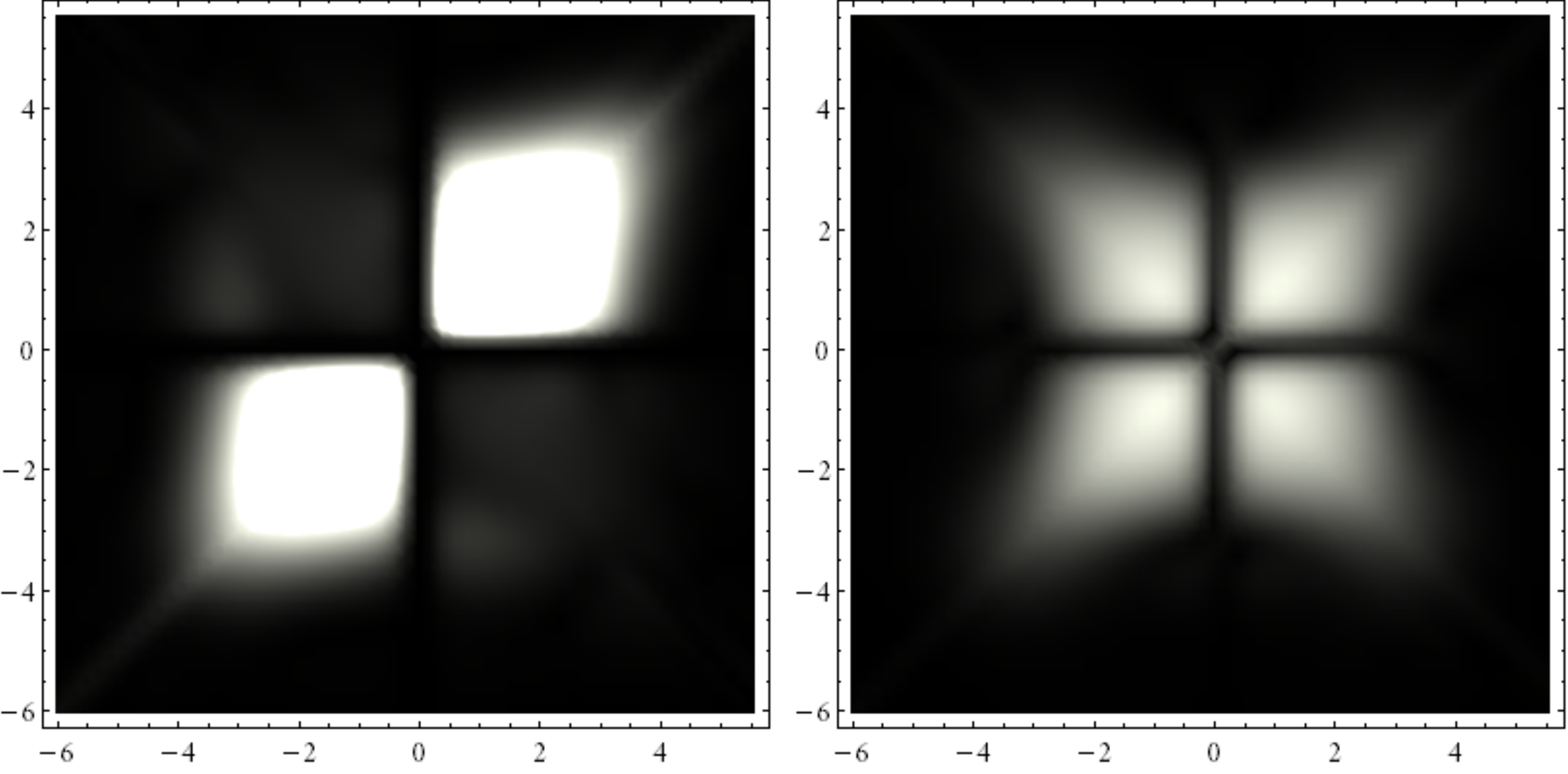}
\caption{The revival of coherence. The coarse grained density matrix after complete splitting (left) and after switching off potential barrier (right). The time of coarse graining is the same: 1.38 s.}
\label{revival}
\end{figure}

The next numerical experiment consists of a sequence of growing barrier, long evolution of the split gas, then rapid decrease of the barrier allowing for the recombination of the gas and finally the coarse graining with the integration time as long as the one enabling the decoherence in the presence of the impenetrable barrier. The coarse grained density matrix is shown in Fig. \ref{revival}. Clearly the mutual coherence is quickly restored. Thus the variance of the population difference is expected to increase to the Poissonian level.

\begin{figure}
\includegraphics[width = 0.4 \textwidth]{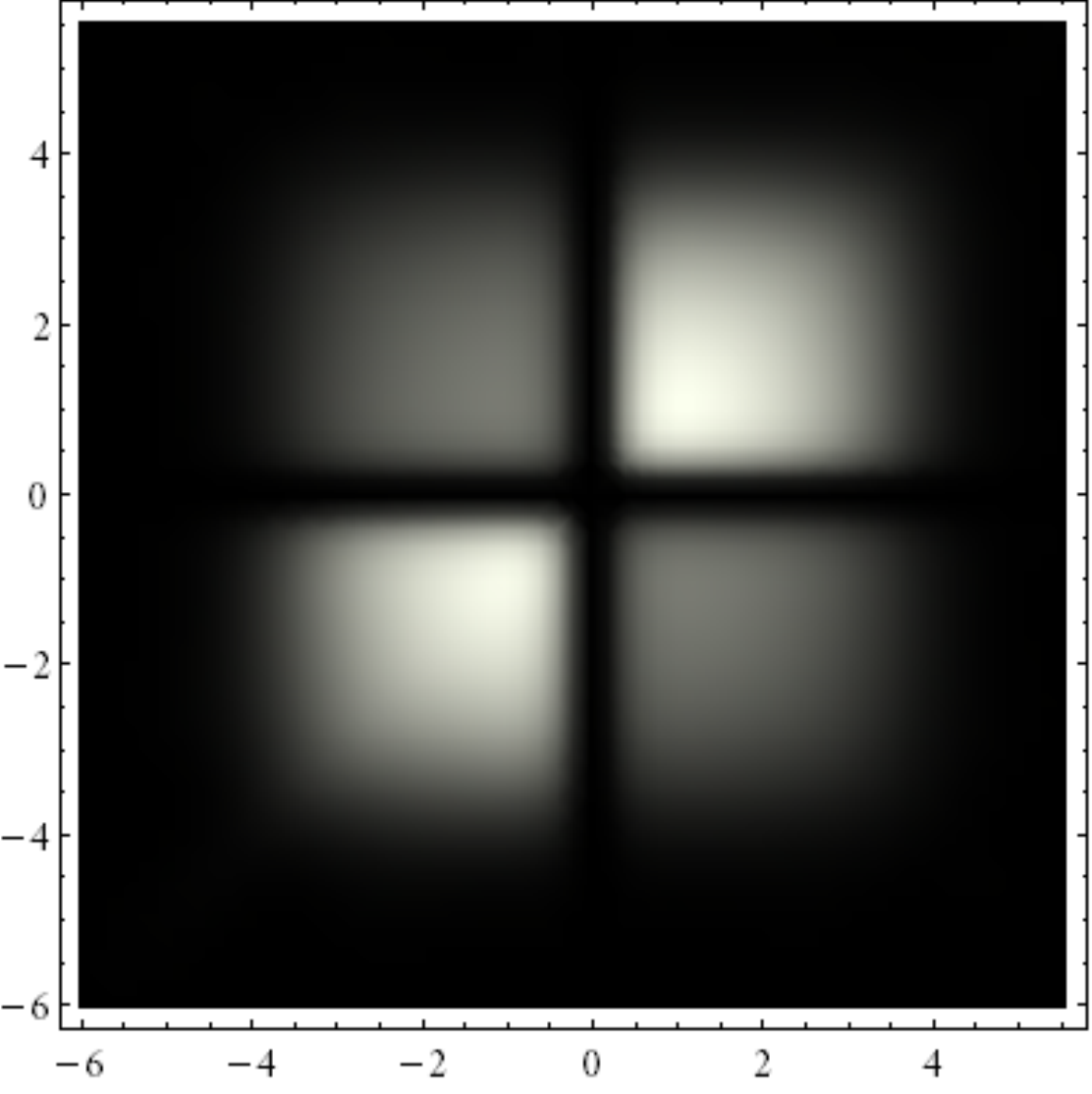}
\includegraphics[width = 0.4 \textwidth]{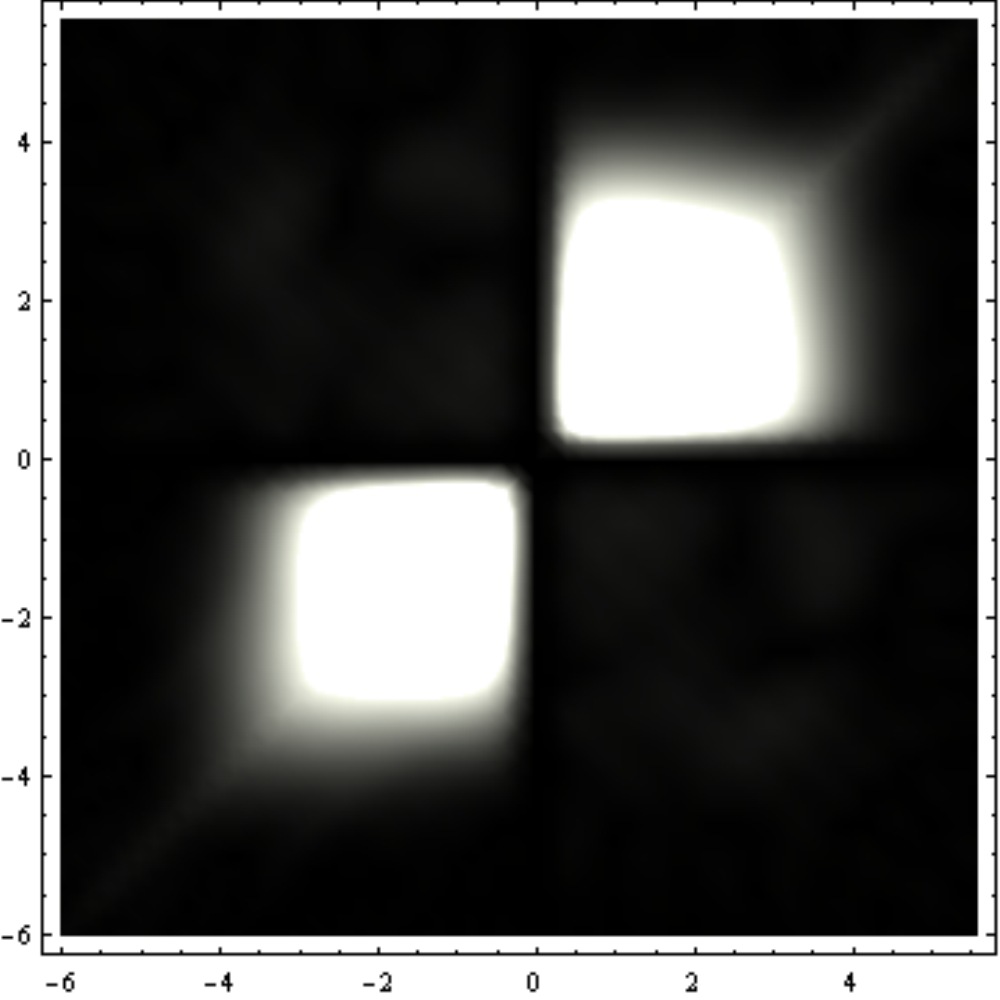}
\caption{The coarse grained density matrices after complete splitting for two different times of splitting: 0.69 s (left) and 0.069 s (right). The time of averaging is the same: 1.38 s.}
\label{fast+slow}
\end{figure}

The speed of raising the barrier matters: as the splitting is done more gently fewer excitations are generated and the system stays closer to its ground state. We illustrate this property in Fig. \ref{fast+slow}, where we compare the coarse grained one-particle density matrix for two speeds of splitting. 

\begin{figure}[h]
\includegraphics[width=1\textwidth]{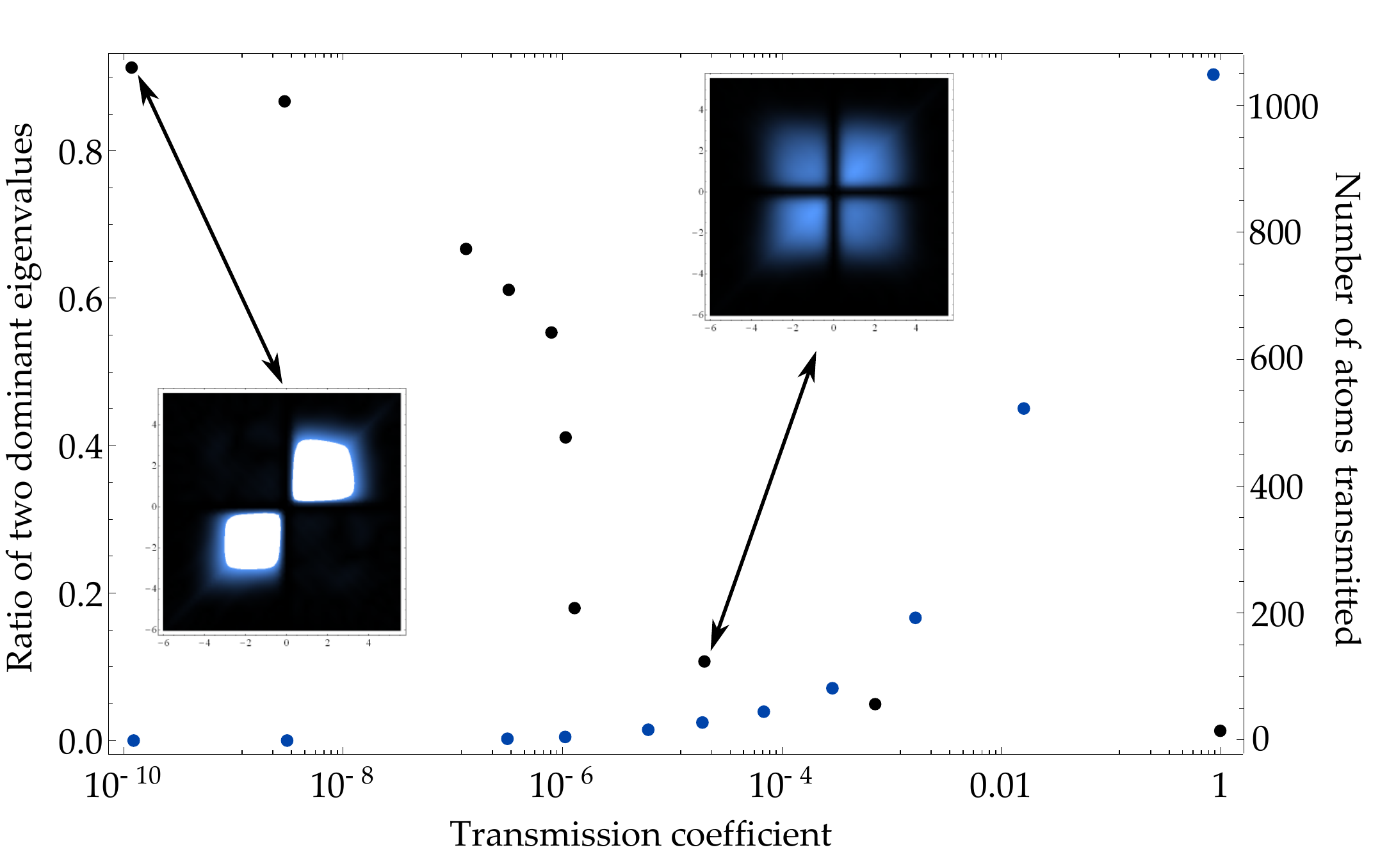}

\caption{The ratio of two dominant eigenvalues of the coarse grained density matrix (black filled circles) and the number of atoms transmitted (red filled circles) as a function of the transmission coefficient of the final barrier.}
\label{evlp_tcf}
\end{figure}

As a next example, consider a barrier which is raised to a level that is not high enough to shutdown the tunnelling. The density matrix, again for the same integration times, is shown in Fig. \ref{evlp_tcf}. When the barrier remains penetrable to allow just a few atoms to switch sides during the coarse graining the coherence is retained..

\begin{figure}[h]
\includegraphics[width=1\textwidth]{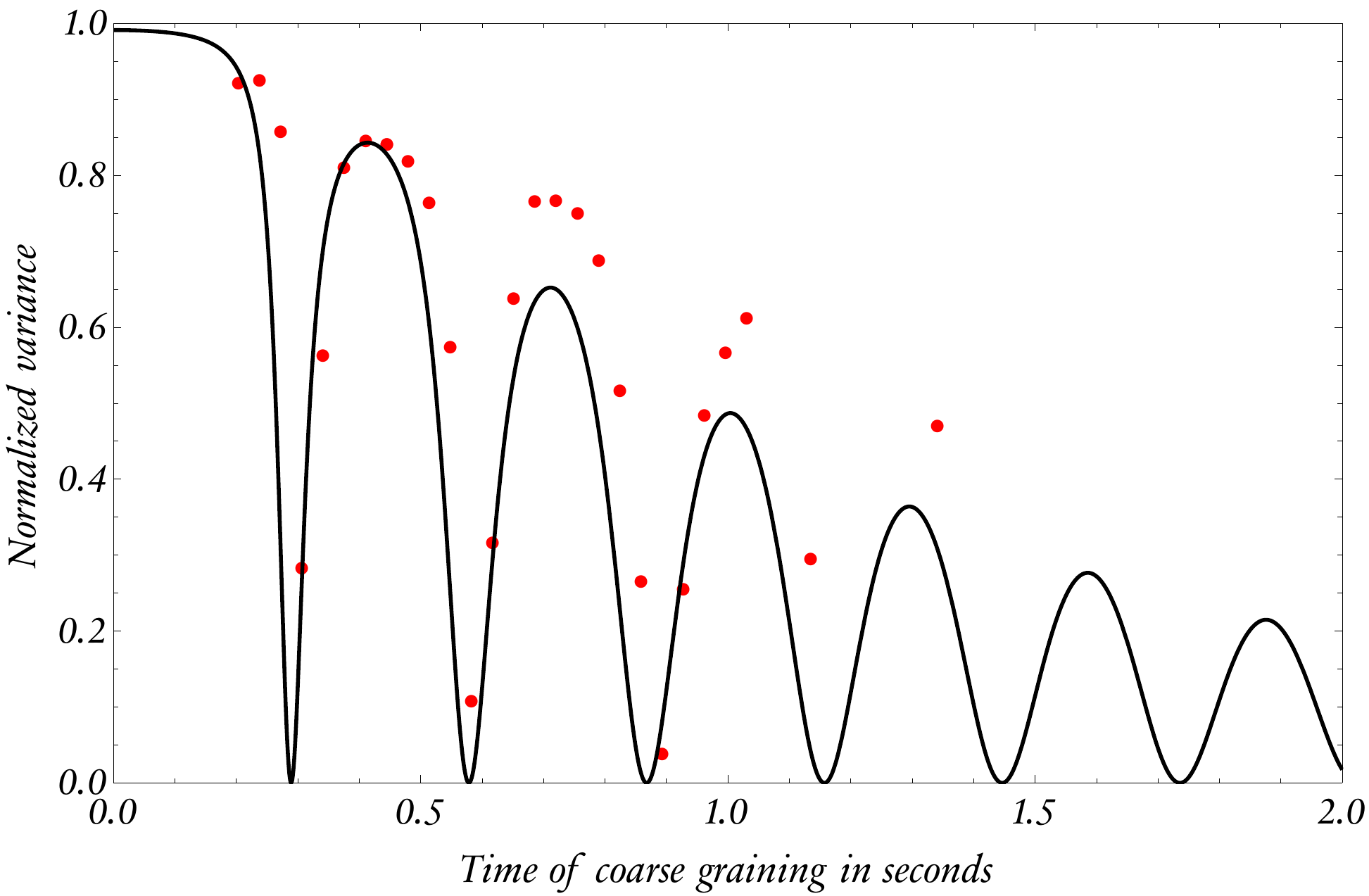}

\caption{The normalized variance as a function of time of coarse graining for interacting gas at temperature $0.16 \ T_c$ (red filled circles). The prediction from a two mode model (black line).}
\label{var_tc_T10}
\end{figure}

\begin{figure}[h]
\includegraphics[width=1\textwidth]{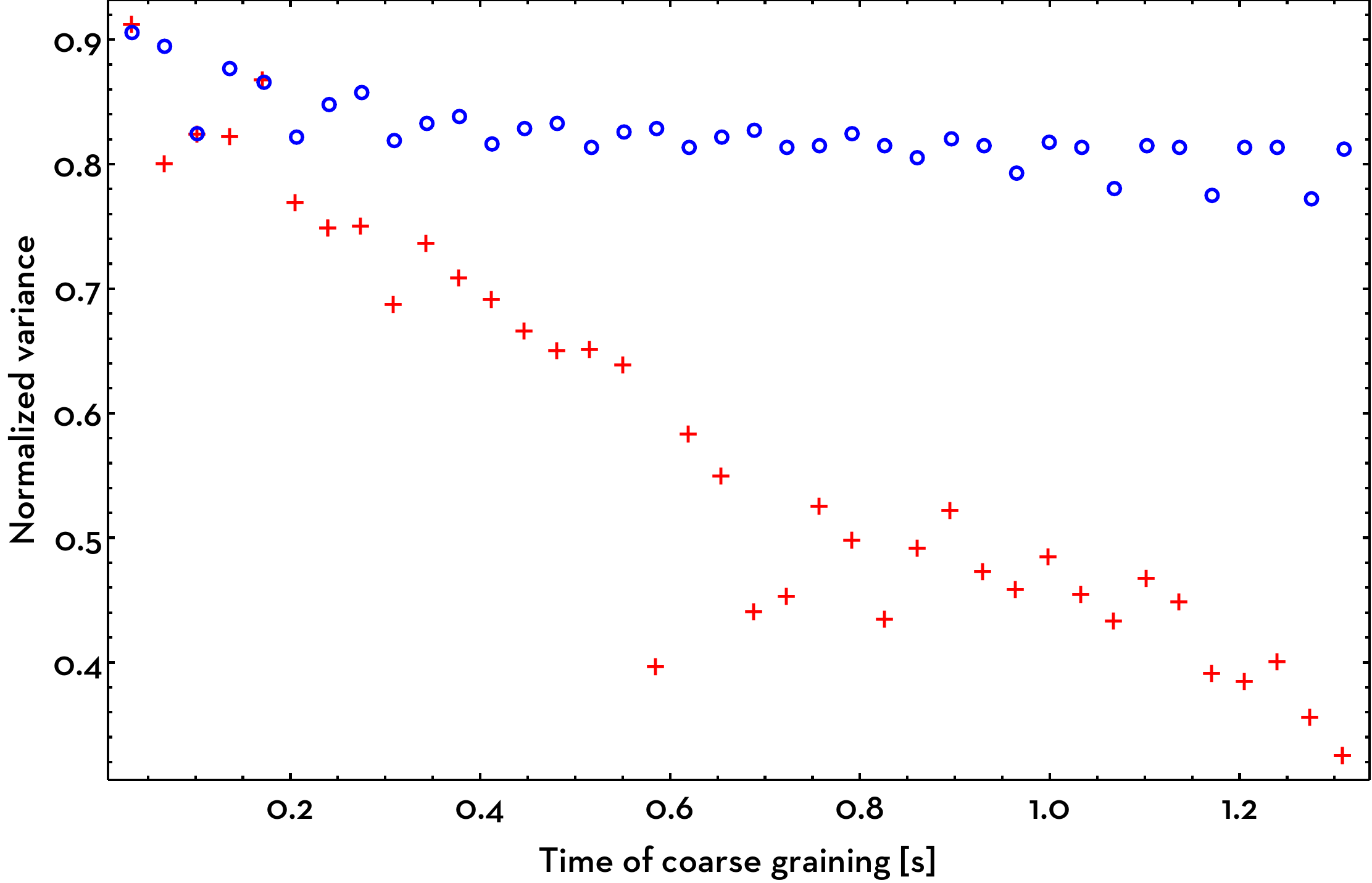}

\caption{The normalized variance as a function of time of coarse graining for interacting gas (red crosses) and for ideal gas (blue open circles) at temperature $0.76 \ T_c$.}
\label{var_tc_T100}
\end{figure}

Finally, we can calculate the variance of atom number difference for different imaging times for low temperature (Fig. \ref{var_tc_T10}) and for high temperature (Fig. \ref{var_tc_T100}).

At low temperature the dependence of variance on time of coarse graining is non-monotonic. The solid line is a result for our naive two mode model.
The characteristic time of the decay of coherence is the inverse of the difference of the left and right chemical potential, hence the the times are shorter for stronger interactions
Note the large difference between the ideal and the interacting gas in Fig. \ref{var_tc_T100}.
The steady decay of the coherence is now due to the presence of many independent highly populated modes with their random phases.

\begin{figure}[h]
\includegraphics[width=1\textwidth]{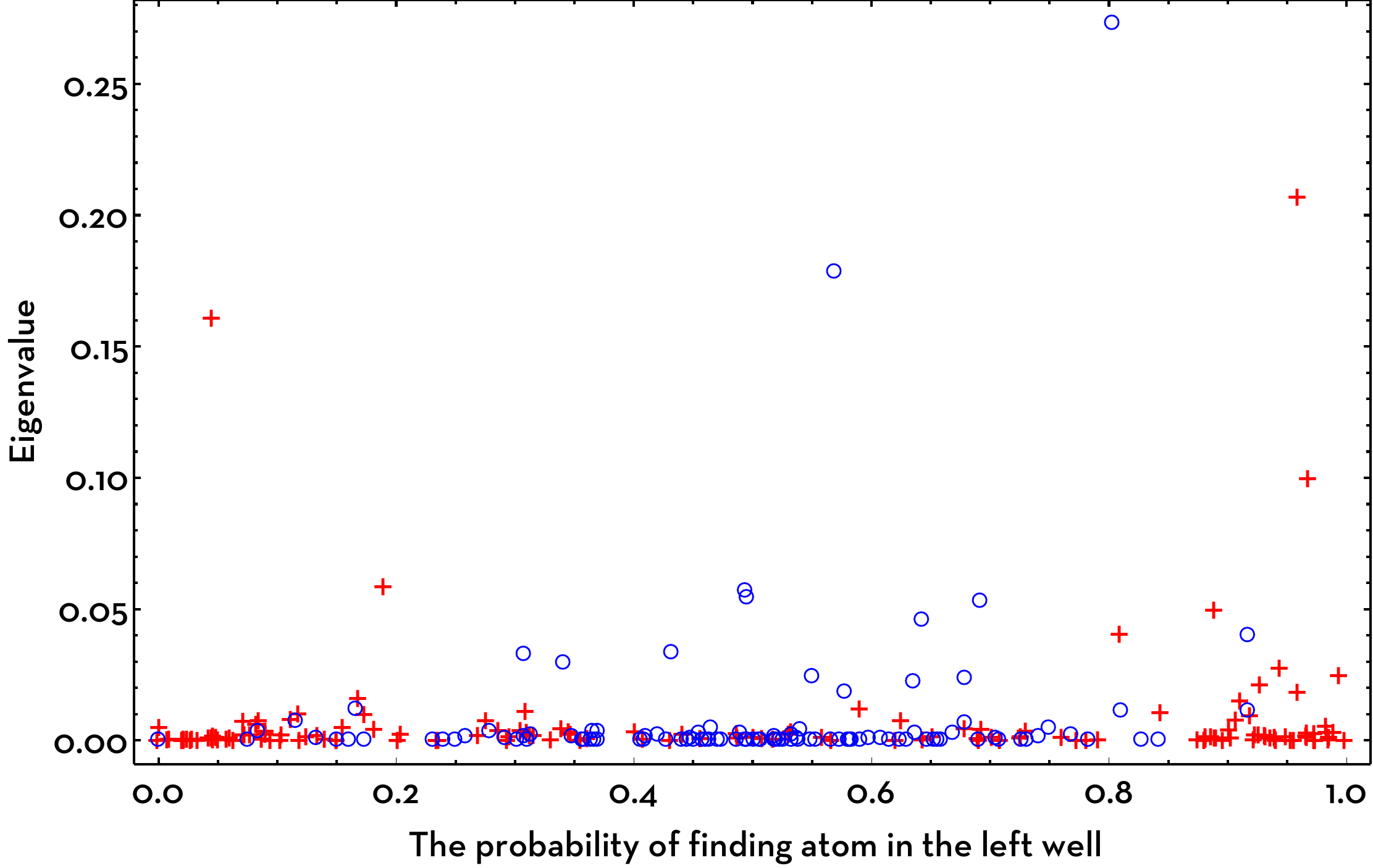}

\caption{The eigenvalues of coarse grained density matrix as a function of probability of finding atoms in the left well for corresponding orbitals for the ideal gas (blue open circles) and for the interacting gas (red crosses) at temperature $0.76 \ T_c$}
\label{p_evl}
\end{figure}

We also plot probabilities $p_L$ of finding atoms in the left well for different eigenvectors as a function of the eigenvalue, Fig. \ref{p_evl}. We again clearly see the difference between the ideal and the interacting gas: for interacting gas for sufficient coarse-graining time most orbitals are localized in the left or in the right well.

The physical separation of the system combined with the contact interaction and with the account for the imperfect observation procedure produce degrading of the coherence. A natural question arises: what happens in the presence of long range forces. To answer this question we have also performed a simulation with a model Gaussian interaction potential with its range significantly longer than the width of the barrier. So even after splitting, the atoms from both wells continue to interact. The decoherence of split condensate was observed again. This leads us to a conclusion that while a nonlinearity in the evolution equation is essential, detailed properties of the interparticle interaction potential are not relevant for our result.

\newpage

\section{Conclusions}

We have shown that the results of a conventional quantum measurement of the Bose-Einstein condensate may, in some cases, depend on the resolution of the measurement procedure, at least according to the classical fields approximation. It would be very interesting to verify this main conclusion of the present paper. Thus changing the duration of the optical pulse and looking at possible modification of the statistics.

We note that our observation times are unrealistically long (around one second). This is probably a result of a strict one-dimensional model for which the Gross-Pitaevskii equation is very close to a fully integrable nonlinear Schr\"odinger equation \cite{zakharov73} thus making thermalization times very long.

The authors acknowledge enlightening discussions with M. Raymer, I. Bloch and T. Pfau. This research was supported by: Polish National Research Center under contract: DEC-2012/04/A/ST2/00090 and by contract research International Spitzenforschung II-2 of the Baden-W\"urtemberg Stiftung, "Decoherence in long range interacting quantum systems and devices".


\clearpage

\begin{thebibliography}{99}

\bibitem{zurek81} W. H. \.Zurek, Phys. Rev. D 24, 1516
(1981).

\bibitem{zurek00} D. A. R. Dalvit, J. Dziarmaga, and W. H. \.Zurek, Phys. Rev. A 62, 013607 (2000).

\bibitem{zurek12} J. Dziarmaga, W. H. \.Zurek, and M. Zwolak, Nature Physics 8, 49–53 (2012).

\bibitem{shin04} Y. Shin, M. Saba, T. A. Pasquini, W. Ketterle, D. E. Pritchard, and A. E. Leanhardt, Phys. Rev. Lett. 92, 050405 (2004).

\bibitem{schumm05} T. Schumm, S. Hofferberth, L. M. Anderson, S. Wildermuth, S. Groth, I. Bar-Joeseph, J. Schmiedmayer, and P. Kr\"uger, Nature Physics 1, 57 (2005).

\bibitem{jo07} G.-B. Jo, Y. Shin, S. Will, T. A. Pasquini, M. Saba, W. Ketterle, D. E. Pritchard, M. Vengalattore, and M. Prentiss, Phys. Rev. Lett. 98, 030407 (2007).

\bibitem{mebrahtu06}A. Mebrahtu, A. Sanpera, and M. Lewenstein, Phys. Rev. A 73, 033601 (2006).

\bibitem{maussang10} K. Maussang, G. E. Marti, T. Schneider, P. Treutlein, Yun Li, A. Sinatra, R. Long, J. Est\`{e}ve and J. Reichel, Phys. Rev. Lett. 105, 080403 (2010).

\bibitem{ketterle97} M. R. Andrews, C. G. Townsend, H.-J. Miesner, D. S. Durfee, D. M. Kurn, and W. Ketterle,
Science 31 January 1997: 275 (5300), 637-641.

\bibitem{kagan97} Yu. Kagan, B. V. Svistunov, Phys. Rev. Lett. 79 3331-3334 (1997).

\bibitem{goral01} K. G\'oral, M. Gajda, and K. Rz\k{a}\.zewski, Opt. Express 8, 92-98 (2001).

\bibitem{sinatra01} A. Sinatra, C. Lobo, and Y. Castin, Phys. Rev. Lett. 87, 210404 (2001).

\bibitem{davis01} M. J. Davis, S. A. Morgan, and K. Burnett, Phys. Rev. Lett. 87, 160402 (2001).

\bibitem{witkowska09} E. Witkowska, M. Gajda, and K. Rz\k{a}\.zewski, Phys. Rev. A 79, 033631 (2009).

\bibitem{metropolis53} N. Metropolis, A. W. Rosenbluth, M. N. Rosenbluth, A. H. Teller, and E. Teller,
The Journal of Chemical Physics 21(6): 1087-1092 (1953).

\bibitem{karpiuk12} T. Karpiuk, P. Deuar, P. Bienias, E. Witkowska, K. Pawlowski, M. Gajda, K. Rz\k{a}\.zewski, M. Brewczyk, Phys. Rev. Lett. 109, 205302 (2012).

\bibitem{goral02} K. G\'oral, M. Gajda, and Kazimierz Rz\k{a}\.zewski, Phys. Rev. A 66, 051602(R) (2002).

\bibitem{brewczyk13} For a review see:  M. Brewczyk, M. Gajda, and K. Rz\k{a}\.zewski, A Classical Field Approach for Bose Gases, p.191 in Quantum Gases: Finite Temperature and Nonequilibrium Dynamics, (ed. N. Proukakis, S. Gardiner, M. Davies, and M. Szymańska) Imperial College Press (2013).

\bibitem{mandel95} L. Mandel and E. Wolf, Optical Coherence and Quantum Optics, Cambridge University
Press (1995).

\bibitem{lodwin55}
P.-O. L\"owdin, Phys. Rev. 97, 1474 (1955).

\bibitem{penrose56} O. Penrose and L. Onsager, Phys. Rev. 104, 576–584 (1956).

\bibitem{zakharov73} V. E. Zakharov and A. B. Shabat, Zh. Eksp. Teor. Fiz. 64, 1627 (1973), Sov. Phys. JETP 37, 823 (1973).

\end{thebibliography}
\end{document}